\documentclass[aps,prl,twocolumn,groupedaddress]{revtex4-1}

\usepackage{graphicx}
\usepackage{color}
\usepackage{comment}
\usepackage{amsmath}
\newcommand{\rd}{\ensuremath{\mathrm{d}}}
\newcommand{\e}{\ensuremath{e}}
\newcommand{\pd}{\ensuremath{\partial}}
\newcommand{\kbt}{\ensuremath{k_\mathrm{B}T}}
\renewcommand{\vec}[1]{\mathbf{#1}}
\begin{document}

%\preprint{}

\title{Quantum symmetry from enhanced sampling methods}


\author{J. Runeson}
\author{M. Nava}
\email{mark.nava@gmail.com}
\author{M. Parrinello}
\affiliation{Department of Chemistry and Applied Biosciences, ETH Z\"{u}rich, and Facolt\`a di Informatica, Istituto di Scienze Computazionali, Universit\`a della Svizzera Italiana, Via G. Buffi 13, 6900 Lugano Switzerland}

\date{\today}

\begin{abstract}
We address the problem of the minus sign sampling for two electron systems using the path integral approach. We show that this problem can be reexpressed as one of computing free energy differences and sampling the tails of statistical distributions. Using Metadynamics, a realistic problem like that of two electrons confined in a quantum dot can be solved. We believe that this is a strategy that can possibly be extended to more complex systems.
\end{abstract}

%\pacs{05.30.-d,05.30.Fk,05.30.Jp}

\maketitle

\section{Introduction}
The numerical simulation of many-body fermion systems is one of the frontiers of contemporary chemistry and physics.
Over the years the quantum chemistry community has developed sophisticated methods to evaluate the ground state properties of
atoms and molecules \cite{helgaker}. Unfortunately these methods scale poorly with the system size.
For this reason a variety of methods have been proposed that transform the calculation of the ground state properties
into one of statistical sampling \cite{releasenode,ceprev,alavi,sorella,pigsnava}.
These approaches have in principle a more benign scaling and lead to exact results within statistical uncertainty.

A severe difficulty is posed by the fact that for fermions the statistical distribution that needs to be sampled is not positive definite. This
is at the heart of the fermionic negative sign problem \cite{signproblem}. In order to circumvent this problem several approaches have been suggested, but so far
only relatively small size systems have been studied. Similarly challenging is the study of Fermi systems at finite temperature \cite{cep_book1}, a problem to which less attention has been devoted.
%If we move from the study of the ground state to finite temperature Fermi systems only a limited number of studies can be reported.

\begin{figure}
\includegraphics[width=\columnwidth]{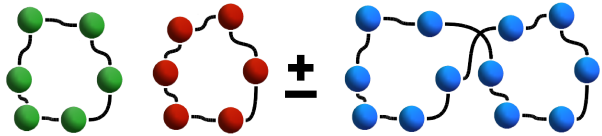}
\caption{Schematic visualization of two particles in the path integral-description. Each particle consists of a set of a necklace of beads connected by harmonic interactions. To describe indistinguishable particles one needs to include contributions from connected necklaces, using positive sign for bosons and negative signs from fermions. \label{fig_beads}}
\end{figure}

Here we limit ourselves to study two-electron systems in order to illustrate the sampling problem and indicate a possible line of attack for more complex systems. Our approach is based on Feynman's path integral
representation of quantum statistical mechanics \cite{feynhibbs}. Neglecting exchange this leads to the well known isomorphism in which each quantum particle
is mapped into a closed necklace of $P$ beads (see Fig.~\ref{fig_beads}a) linked by harmonic forces and interacting via suitably scaled-down
potentials. This isomorphism has been successfully used to solve problems where exchange can be neglected \cite{kalos,fcenter,pimd_par,berne}.

However, exchange effects complicates this picture. In fact, due to the indistinguishablility of
particles, processes in which two necklaces merge into a single one need to be considered (see Fig.~\ref{fig_beads}b) \cite{wolchand}. For bosons these processes are
added with a plus sign, while for fermions they are to be added with a negative sign. In this latter case the distribution that needs to be sampled is no
longer positive definite.

In this paper we show that the problem of sampling a system of two fermions is related to a calculation of the free energy difference between two classical systems, one
in which the particles are distinguishable and one in which, due to exchange, the two polymers have merged into a single one (Fig.~\ref{fig_beads}b).

This calculation is rather delicate since there are large statistical errors and a careful sampling of the tails of the distributions is needed for an accurate result. To this effect we use Metadynamics \cite{meta_pnas,wtm,valsson},
an enhanced sampling method developed in our group. After having established the strategy we apply it to the case of two non-interacting fermions in a harmonic well,
a problem that can be analytically solved and that clearly illustrates the nature of the sampling problem. We then switch our attention to the realistic case of two electrons in a quantum dot,
a system that can be realized in the laboratory \cite{qdot3}.
Our results are in good agreement with exact calculations.

%Quantum indistinguishability can be represented in the PI isomorphism as an exchange between any two beads with the same index but belonging to different atoms, this amounts at reconnecting the harmonic
%interactions as depicted in Figure \ref{fig_beads} for $N=2$ particles. This can happen for any bead index $m$ and every time this is done the configuration flips from a configuration where the two ring polymers
%are joined into a single one consisting of $2P$ beads (Figure \ref{fig_beads}b) to one where the two particles are represented by two different polymers of size $P$ (Figure \ref{fig_beads}a) and vice-versa. In the configurations
%where the polymers are joined the two quantum particles can not be distinguished, and since to go back to the other topology another exchange could happen anywhere in the polymer the effect is that every bead loses its association with
%its former quantum particle, as should be expected since the particles are now indistinguishable.

%%%%%%%%%%%%%%%%%%%%%%%%%%%%%%%%%%%%%%%%%%%%%
\section{Method}
We start by considering two distinguishable particles. The path integral representation of their partition function is

%for which the effective potential in the PIMD description is $V_{oo}$ (the subscript is used to remind the reader of the two-ring
%structure in Figure~\ref{fig_beads}a). In such a case the partition function can be written

\begin{eqnarray}
Z_{oo} = \int \rd R_1 \rd R_2 \: \e^{-\beta V_{oo}(R_1,R_2)}  \label{Zdist}
\end{eqnarray}
where $\beta$ is the inverse temperature, $R_1=(\vec{r}_1^{\,1},...,\vec{r}_1^{\,P})$ is the set of coordinates representing the $P$ beads of particle $1$ and $R_2$ is defined analogously.
The effective potential $V_{oo}$ describes two necklaces (as displayed in Fig.~\ref{fig_beads}a) with a rescaled interaction potential $V(\vec{r}_1,\vec{r}_2)/P$,
\[ V_{oo} =  \sum_{i=1}^P\left( \sum_{n=1}^2 \frac{mP}{2\hbar^2\beta^2}(\vec{r}_n^{\,i+1}-\vec{r}_n^{\,i})^2 + \frac{1}{P}V(\vec{r}_1^{\,i},\vec{r}_2^{\,i})\right), \]
where the bead index $i$ is cyclic, $\vec{r}_n^{\,P+1}\equiv \vec{r}_n^{\,1}$.
In order to describe indistinguishable particles, also configurations of the connected polymer-type (Fig.~\ref{fig_beads}b), with effective potential
\begin{equation} 
V_{O} = \sum_{j=1}^{2P}  \frac{mP}{2\hbar^2\beta^2}(\vec{r}^{\,j+1}-\vec{r}^{\,j})^2 + \frac{1}{P}\sum_{i=1}^P V(\vec{r}_1^{\,i},\vec{r}_2^{\,i}), \label{VO}
\end{equation}
need to be taken into account. Here and in the following the subscripts $oo$ and $O$ are meant to remind one of the different topologies of the necklaces. In the first term of Eq.~(\ref{VO}) we have arranged the necklace bead coordinates $\vec{r}_1^{\,i}$ and $\vec{r}_2^{\,i}$ to form a single vector $\vec{r}^j$ of dimension $2P$.

The partition function for indistinguishable particles can be written
\begin{eqnarray}
Z_{I} = \frac{1}{2}\int \rd R_1 \rd R_2 \: \left( \e^{-\beta V_{oo}} \pm \e^{-\beta V_{O}} \right)  \label{Zindist}
\end{eqnarray}
where $I=B,F$ labels the boson (plus sign) or the fermion state (minus sign). In the latter case, the integrand is not positive definite, making sampling difficult. Instead we write Eq.~(\ref{Zindist}) as
\begin{eqnarray} \label{ZW}
Z_{I} = \frac{1}{2} \int \rd R_1 \rd R_2 \: \e^{-\beta V_{oo}} \: W_{I}  \\
W_{I} = 1\pm\e^{-\beta\left(V_O-V_{oo}\right)}.
\end{eqnarray}
The thermal average of an operator $\hat{O}$ local in coordinate representation can be written
\begin{eqnarray}
\langle \hat{O} \rangle = \frac{\langle O W_{I} \rangle_{oo} }{\langle W_{I} \rangle_{oo}} \label{wavg}
\end{eqnarray}

\begin{figure}
\includegraphics[width=\columnwidth]{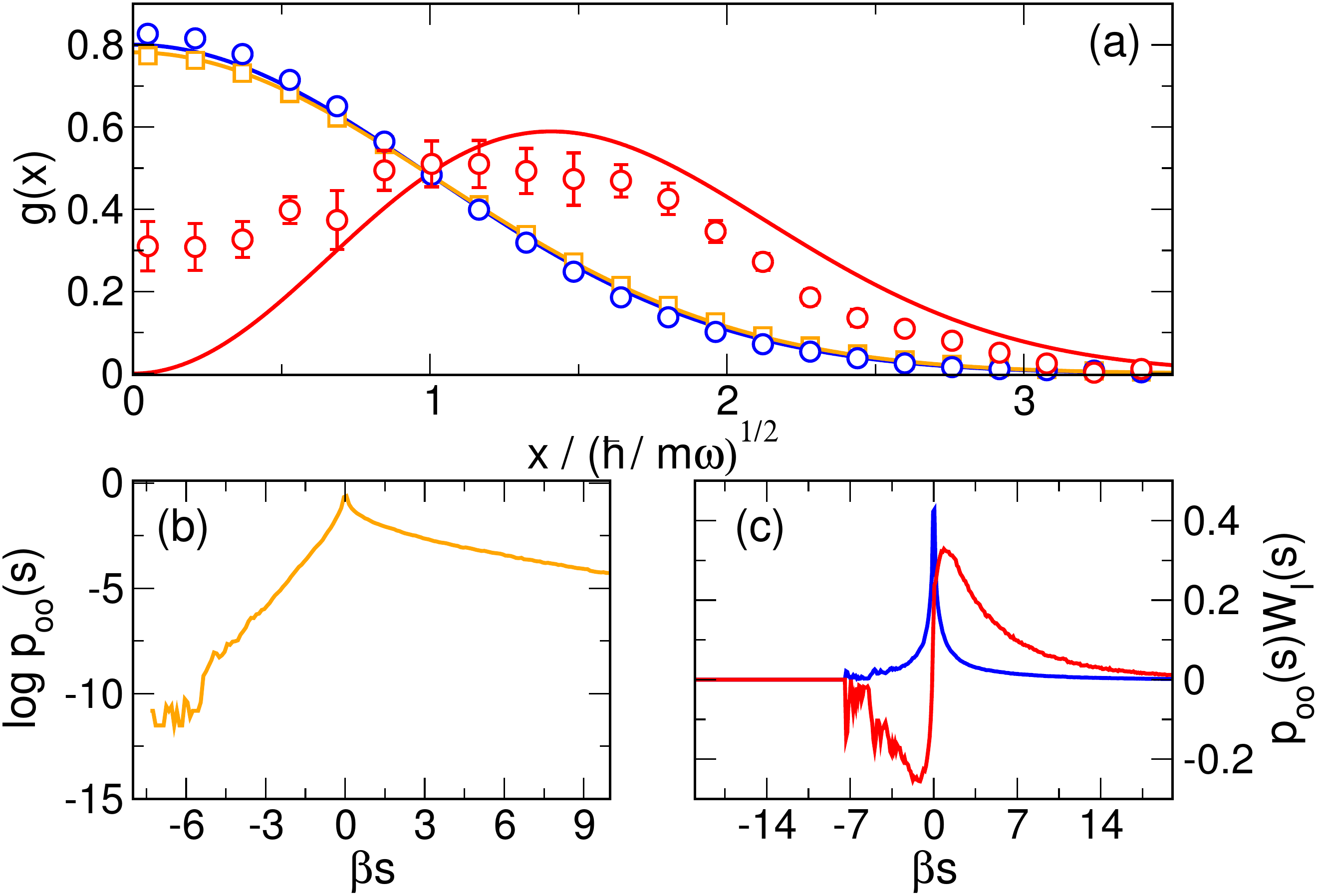}
\caption{Non-biased simulation of the indistinguishable particles in a harmonic well: (a) pair distribution function for bosons (blue) and fermions (red). Our results are denoted by cicles, the analytical results are the continuous lines. The squares are the result of a simulation in which exchange is ignored.
(b) Logarithm of the $p_{oo}(s)$ distribution.
(c) Plot of the integrand in $\langle W_I\rangle_{oo}=\int p_{oo}(s)W_I(s)\rd s$ (the height of the peaks is irrelevant and has been rescaled for visualization purposes).
The simulation has been conducted in the quantum regime $(\beta\hbar\omega=3)$.
\label{fig_nometa}}
\end{figure}

\noindent
where $\langle...\rangle_{oo}$ denotes an average over a system of distinguishable particles.
This means that also bosons and fermions can be described through a simulation of distinguishable particles,
if one takes the quantum symmetry into account with the weight $W_{I}(s) = 1 \pm \e^{-\beta s}$,
where $s\equiv V_O - V_{oo}$ is the difference in spring energy between the two ring polymer topologies.
This procedure is analogous to other approaches where one drives the simulation with a positive definite distribution and corrects it with the sign function \cite{ceperley1995fermion}.
Here the sign function is replaced by $W_{I}(s)$. The average of $W_I(s)$ can be rewritten as $\langle W_I\rangle_{oo} = \int \rd s\, p_{oo}(s)W_I(s)$, where $p_{oo}(s)$ is the probability distribution of the variable $s$ in the distinguishable particle ensemble.

When quantum exchange effects become important, the overlap between the ensemble distributions of distinguishable and indistinguishable particles can be very small. In particular for fermions the statistical distribution that needs to be sampled has large contributions coming from configurations that are rarely sampled. 
 This becomes evident if one attempts at simulating a system as simple as that of two non-interacting quantum particles in a harmonic well.

In the following we shall use path integral molecular dynamics (PIMD) to sample these distributions, however what we describe below is also fully compatible with Monte Carlo sampling.
Using Eq.~(\ref{wavg}) we calculate the pair distribution function (Fig.~\ref{fig_nometa}). One can contrast the boson case, where only small deviations at short distances can be seen, with that of the fermions, where one fails to reproduce the exchange hole.
The reason for this behaviour can be understood if we contrast $p_{oo}(s)W_B(s)$ with $p_{oo}(s)W_F(s)$. In the first case most of the contributions come from the $s=0$ region, while for fermions $\langle W_F\rangle$ results from a delicate cancellation between positive and negative contributions, most of which come from the poorly sampled $s<0$ region.
Thus the fermionic sign problem can be expressed as a problem of sampling the tail of a probability distribution.

A way of enhancing the fluctuations of a given variable in a controlled manner is offered by Metadynamics \cite{meta_pnas,wtm}. To make connection with Metadynamics literature we shall use $s$ defined above as collective variable. Following the Metadynamics prescription we add to the Hamiltonian a time-dependent term of type
\[ V(s,t)=\sum_{t'<t}w(t')\exp\left(-\frac{\left[s-s(t')\right]^2}{2\sigma_G^2}\right) \]
where the height of the Gaussian of width $\sigma_G$ is given by the well-tempered prescription
\[ w(t)=w(0)\exp\left(-\frac{1}{\gamma - 1}\beta V(s,t)\right) \]
and the Gaussians are added at a specified time interval $\tau_G$. This procedure has been proven to be rigorous \cite{WT_Dama} and leads asymptotically to sampling the distribution 
\[p_V(s) = p(s)^{1/\gamma} \]
where $p(s)$ is the distribution in the unbiased run, while $p_V(s)$ is the one sampled in the Metadynamics run.
A simple reweighting procedure allows to calculate equilibrium expectation values of any operator from the Metadynamics run \cite{reweighting}.

%This can be solved by using enhanced sampling techniques, like for example Metadynamics\cite{meta_pnas,wtm}, where the fluctuations of a collective variable (CV) $s$ are enhanced
%by adding to the Hamiltonian a time-dependent bias potential
%\begin{eqnarray}
%\[ V(s,t)=\sum_{t'<t}w(t')\exp\left(-\frac{\left(s-s(t')\right)^2}{2\sigma_G^2}\right) \]
%\end{eqnarray}
%We use the energy difference $s=V_O-V_{oo}$ as our CV.
%In practice the bias is added by depositing Gaussians of width $\sigma_G$ at times spaced by a fixed interval $\tau_G $. The height $w(t)$ of the Gaussian is determined by the well-tempered equation\cite{wtm}:
%\begin{equation}
%\[ w(t)=w(0)\exp\left(-\frac{V(s,t)}{k_B\Delta T}\right) \]
%\end{equation}
%where $w(0)$ is the initial Gaussian height, and $\Delta T$ a temperature. %This procedure converges\cite{WT_Dama} modulo an irrelevant constant to $F(s)$.
%The effect of the bias on the thermal averages can then be removed on the fly with a reweighting procedure\cite{reweighting}. This procedure has been shown to be rigorous\cite{WT_Dama}.

\begin{figure}
\includegraphics[width=\columnwidth]{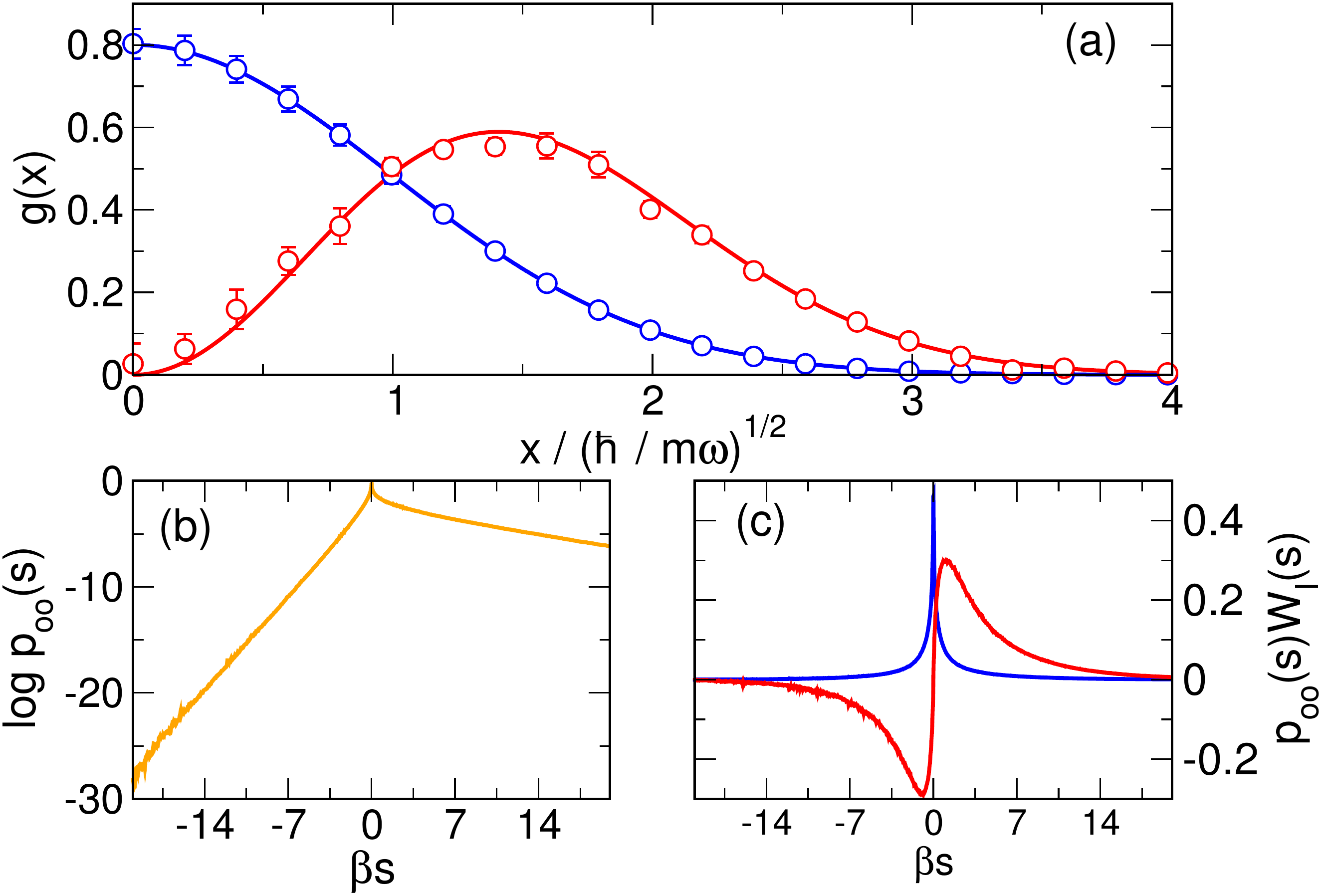}
\caption{Results of a Metadynamics simulation for two particles in a harmonic potential at $\beta\hbar\omega=3$. For a definition of the symbols, see Fig.~\ref{fig_nometa}. \label{fig_meta}}
\end{figure}

The results obtained using Metadynamics are shown in Fig. \ref{fig_meta}. Compared to the results in Fig. \ref{fig_nometa}, the probability distribution $p_{oo}(s)$ is now sampled efficiently
down to large negative values of $s$. As a consequence the negative domain of the integrand of Eq. \ref{Zindist} for the fermionic system is well resolved and the
cancellation between positive and negative regions can be brought under control.  The pair distributions are now in fact close to the theoretical curves; the improvement is especially clear for the
Fermi particles, where the strong repulsion at short distances that comes from the Pauli exclusion principle is accurately represented.

The results shown in Fig.~\ref{fig_nometa}--\ref{fig_meta} were obtained for a system in the quantum regime ($\beta\hbar\omega=3$) and the number of beads in the isomorphism was $P=10$. The equations of motion were integrated using a time step of 0.023/$\omega$ and a colored-noise thermostat \cite{GLE} purposely designed for path integral simulations was employed. The simulations were run for $2\times 10^7$ time steps. In the Metadynamics runs, Gaussians of initial height $w(0)=0.5\kbt$, width $\sigma_G=4\kbt$ and $\gamma=4$ were deposited every $\tau_G=2000$ steps.

%The parameter settings used to obtain Figs.~\ref{fig_nometa}--\ref{fig_meta} was a timestep of $0.023/\omega$, $\hbar\omega=3\kbt$, $P=10$ and $2\times 10^{7}$ steps. To ensure a canonical distribution we used a colored-noise thermostat\cite{GLE} purposely designed for path integral simulations. The same setup was used in the Metadynamics run, where Gaussians with $\sigma_G=4\kbt$, $w(0)=0.5\kbt$ and $\gamma=4$ where deposited every 2000 steps. Sampling started after the Metadynamics bias had been building up for $2\times 10^{8}$ steps.

The calculation of energies, in particular quantum kinetic energies, poses some well known problems that we solve by applying the virial theorem \cite{berne2}, obtaining the energy estimator
\begin{align}
\begin{split}
 &\langle E \rangle =\frac{1}{\langle W_I\rangle_{oo}} \left\langle \frac{1}{P}\sum_i V(\vec{r}_1^{\,i},\vec{r}_2^{\,i}) +  \frac{n_d}{2\beta} \left(2\pm\e^{-\beta\Delta V}\right) \right. \\
 &\left. + \frac{1}{2P}\sum_{i,n} [(1\pm\e^{-\beta\Delta V})\vec{r}^{\,i}_{n} -\bar{\vec{r}}_n - \e^{-\beta\Delta V}\bar{\vec{r}}] \cdot \frac{\pd V}{\pd \vec{r}^{\,i}_{n}}  \right\rangle_{oo} \label{energy_estimator}
 \end{split}
 \end{align}
 where $n_d$ is the number of dimensions, $\bar{\vec{r}}_n$ the centroid of particle $n$ and $\bar{\vec{r}}$ the center of mass of the two particles. 
Alternatively, at low temperatures, once the energy for the boson system is known, the energy of the fermion variant can also be calculated as 
\begin{equation}
 E_F \approx E_B + F_F-F_B \label{efeq}
 \end{equation}
with
%It is interesting to note that also the difference in \emph{free} energy between bosons and fermions
\begin{equation}
F_F-F_B = -\frac{1}{\beta}\ln\frac{Z_F}{Z_B} = -\frac{1}{\beta}\ln\frac{1-\frac{Z_O}{Z_{oo}}}{1-\frac{Z_O}{Z_{oo}}}, \label{bfed}
\end{equation} 
since in the zero-temperature limit free energies are very close to energies.
The ratio $Z_O/Z_{oo}$ can be best estimated by using the Bennett method \cite{bennett1}
%can be calculated once $Z_O/Z_{oo}$ is known. This latter can be estimated using the Bennett method\cite{bennett1}
\begin{equation}
\frac{Z_{O}}{Z_{oo}} = \frac{\left\langle f(\beta s+C)\right\rangle_{oo}}{\left\langle f(-\beta s-C) \right\rangle_{O}}\e^C \label{bennett}
\end{equation}
where $f(x)=(1+\e^{x})^{-1}$ and $C$ is a shift constant that is varied to increase the overlap between the sampling distributions $p_O(s)$ and $p_{oo}(s)$. This observation will be made use of in the following.

\section{Results}
So far we have considered a free particle toy model. We now study the more realistic system
of two interacting electrons in a quantum dot, that has recently received attention because of its possible applications in
semiconductors and electronic devices \cite{qdot1,qdot2,qdot3}.
The electrons are under the influence of a two-dimensional harmonic potential $U(x,y)$ and a rescaled Coulomb interaction $V_{C}(r)$
\begin{eqnarray}
U(x,y) = \frac{1}{2}m^{\star}\left(\omega^2_x x^2 + \omega^2_y y^2\right) \\
V_{C}(r)=\frac{\gamma_C e^2}{4\pi\epsilon_r\epsilon_0 r}
\end{eqnarray}
where $m^\star$ is the effective mass of the electrons, $\omega_x$ and $\omega_y$ regulate the confinement along $x$ and $y$, $\epsilon_r$ is the relative dielectric constant and
$\gamma_C$ is a parameter that takes into account finite size effects on the vanishing $z$ direction of the dot. Following Ref. \cite{qdot3}, we set $m^\star/m_e = 0.07$, $\epsilon_r=12.5$ and $\gamma_C=0.9$, that are appropriate parameters for a real life quantum dot.
The physics of these dots is determined by the anisotropy of the confinement, $\eta = \omega_y/\omega_x$ and the Wigner parameter $R_W = \frac{V_{C}(l_0)}{\hbar\omega_0}$ that is the ratio between
the typical interaction energy and the single particle levels splitting in a averaged confinement $\omega_0=\sqrt{(\omega_x^2+\omega_y^2)/2}$.
The parameter $l_0= \sqrt{\hbar/m^\star\omega_0}$ is the characteristic length of the dot. 
%The strength of the Coulomb interaction is measured by $R_W$: for $R_W = 0$ the quantum dot has the same physics of two non-interacting electrons in a harmonic well.

Neglecting spin-orbit coupling, the spin and orbital part of the wavefunction can be decoupled. In this case two electrons in a quantum dot can be either in a singlet or a triplet state. In the singlet state the spin part is antisymmetric and therefore the orbital part nedds to be symmetric to ensure the antisymmetry of the global wavefunction. In the triplet state the converse is true, namely the spin part is symmetric while the orbital part is antisymmetric.
%The singlet state has the spin part antisymmetric under the exchange of the two electrons and thus the coordinate part of the wavefunction is Bose symmetric, conversely triplet states have a fermionic coordinate part. 

\begin{figure}
\includegraphics[width=\columnwidth]{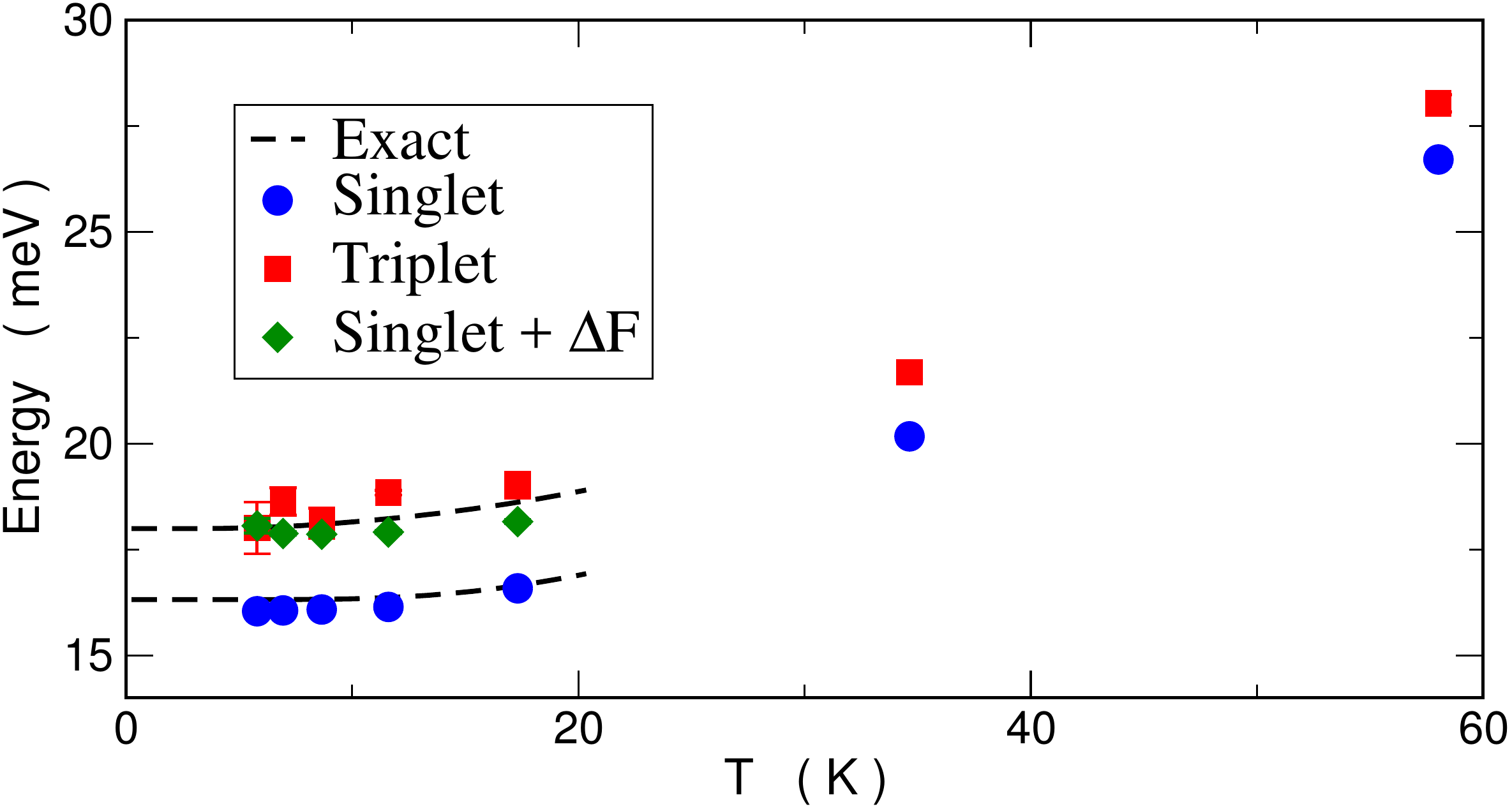}
\caption{Total energy as function of temperature for two electrons in a quantum dot with $\hbar\omega_0=5.1$ meV, $R_W=1.34$ and $\eta=1.38$. Dashed lines are the energies obtained from
exact diagonalization of the Hamiltonian. When not otherwise indicated, the error bars are smaller than the size of the dots. The computation was done from $10^8$ samples with 15 beads at $T=11.6$ K and otherwise such that $\beta/P$ was constant.} \label{fig_energies}
\end{figure}

Using the scheme described above, we calculated the energies of these two states as a function of temperature (see Fig.~\ref{fig_energies}) for realistic values of $R_W$ and $\eta$ \cite{qdot3}. The triplet state energy at low temperature is also calculated using Eq.~(\ref{efeq}). This second estimation leads to smaller statistical errors. 

The parameter settings used in these calculations were a timestep of 1 fs, $\beta/P=0.067$ meV$^{-1}$, $\sigma_G=10\kbt$, $w(0)=0.5\kbt$, $\gamma=6$, $\tau_G=10^4$ steps and the same thermostat as in the free particle system was implemented. In total $5\times 10^8$ MD steps were made and samples where acquired in intervals of 5 steps. In the biased simulation this was preceded by another $5\times 10^8$ steps building up bias.

%The fermionic energies at low temperature are calculated using the estimator in Eq.(\ref{energy_estimator}) or by adding to the bosonic energy the free energy difference computed from Eq.~(\ref{bfed}) and (\ref{bennett}). Note that this second approach leads to smaller statistical errors. However at higher temperature we cannot assume that $F_F-F_B$ is approximately equal to the energy difference.

It is also interesting to study the charge denstity and its variation as a function of the ellipticity parameter $\eta$ contrasting the singlet and the triplet case. In the spherical case $(\eta=1)$ the charge density has a spherical symmetry in both the singlet and the triplet case, however the spread is higher in the triplet. We also plot the quantity $\rho_T-\frac{1}{2}\rho_S$ that in a single particle mean field approximation gives information on the first excited state for low temperatures. As $\eta$ is increased the 2D-rotational symmetry is broken and in the triplet the two electrons tend to be localized at different possitions. This is also reflected in $\rho_T-\frac{1}{2}\rho_S$, that has a nodal plane in the middle.

\begin{figure}
\includegraphics[width=\columnwidth]{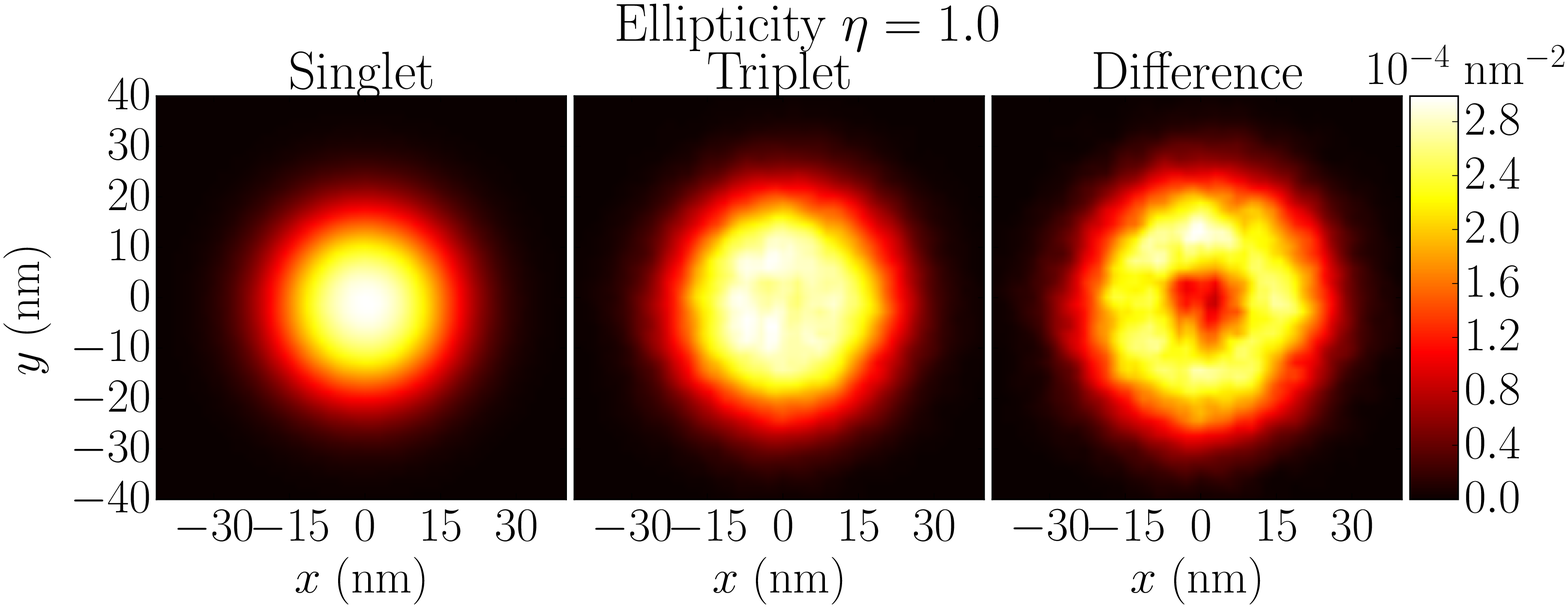}
\includegraphics[width=\columnwidth]{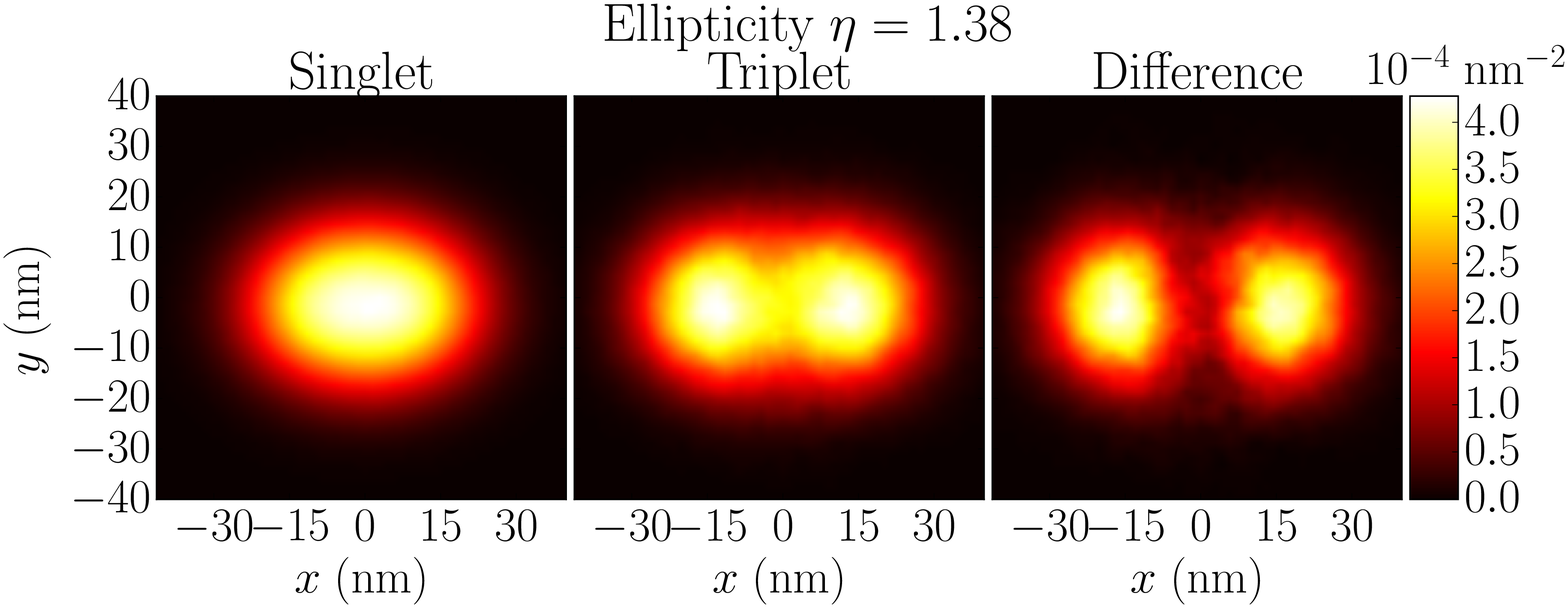}
\caption{ Electron density of singlet and triplet states for a two-particle quantum dot with $R_W=1.34$ for different ellipticities $\eta$ of the confining potential (reported in the figure).
%The ellipticity value, from top to bottom, are $\eta=1$, $\eta=1.4$ and $\eta=3$.
The third column shows the contribution to the triplet density
given by the first excited single-particle state, computed as the difference between the triplet and singlet densities, $\rho_T-\frac{1}{2}\rho_S$. These results where computed from $10^8$ samples at $T=11.6$ K with $P=15$ beads.
\label{fig_densities}}
\end{figure}

\section{Conclusions}
The purpose of this work was to show how the fermionic sign problem can be thought of as a sampling issue that can be solved by enhancing the distribution tails. We have solved a two-body problem but extension to larger systems appear possible and would require separating even and odd permutations. %analyzed in terms of the problem of enhancing the sampling of the tails of a distribution, and then solved -- at least for few-particle systems --
%with an approach that makes us of enhanced sampling methods. Extension to larger systems would require separating the odd and even permutation and use this difference as collective variable.
%This approach is fully compatible with {\it ab-initio} PIMD and can open the way to the study of nuclear quantum exchange effects in chemical compounds. In the process, we have also shown how to obtain free energy differences between states with different quantum symmetry. 

We are very well aware of the limitation of what has been achieved here. It would thus be presumptuous on our side to state that the minus sign problem has been solved. However we have indicated a new way of attacking this long-standing problem. How far this can be carried remains to be seen.
\begin{acknowledgments}
\section{Acknowledgements}
We acknowledge the Swiss National Science Foundation Grant Nr. 200021\_169429/1 and the
European Union Grant Nr. ERC-2014-AdG-670227/VARMET for funding.
\end{acknowledgments}

\bibliography{qsymm}

\providecommand{\noopsort}[1]{}\providecommand{\singleletter}[1]{#1}%
\begin{thebibliography}{26}%
\makeatletter
\providecommand \@ifxundefined [1]{%
 \@ifx{#1\undefined}
}%
\providecommand \@ifnum [1]{%
 \ifnum #1\expandafter \@firstoftwo
 \else \expandafter \@secondoftwo
 \fi
}%
\providecommand \@ifx [1]{%
 \ifx #1\expandafter \@firstoftwo
 \else \expandafter \@secondoftwo
 \fi
}%
\providecommand \natexlab [1]{#1}%
\providecommand \enquote  [1]{``#1''}%
\providecommand \bibnamefont  [1]{#1}%
\providecommand \bibfnamefont [1]{#1}%
\providecommand \citenamefont [1]{#1}%
\providecommand \href@noop [0]{\@secondoftwo}%
\providecommand \href [0]{\begingroup \@sanitize@url \@href}%
\providecommand \@href[1]{\@@startlink{#1}\@@href}%
\providecommand \@@href[1]{\endgroup#1\@@endlink}%
\providecommand \@sanitize@url [0]{\catcode `\\12\catcode `\$12\catcode
  `\&12\catcode `\#12\catcode `\^12\catcode `\_12\catcode `\%12\relax}%
\providecommand \@@startlink[1]{}%
\providecommand \@@endlink[0]{}%
\providecommand \url  [0]{\begingroup\@sanitize@url \@url }%
\providecommand \@url [1]{\endgroup\@href {#1}{\urlprefix }}%
\providecommand \urlprefix  [0]{URL }%
\providecommand \Eprint [0]{\href }%
\providecommand \doibase [0]{http://dx.doi.org/}%
\providecommand \selectlanguage [0]{\@gobble}%
\providecommand \bibinfo  [0]{\@secondoftwo}%
\providecommand \bibfield  [0]{\@secondoftwo}%
\providecommand \translation [1]{[#1]}%
\providecommand \BibitemOpen [0]{}%
\providecommand \bibitemStop [0]{}%
\providecommand \bibitemNoStop [0]{.\EOS\space}%
\providecommand \EOS [0]{\spacefactor3000\relax}%
\providecommand \BibitemShut  [1]{\csname bibitem#1\endcsname}%
\let\auto@bib@innerbib\@empty
%</preamble>
\bibitem [{\citenamefont {Helgaker}\ \emph {et~al.}(2014)\citenamefont
  {Helgaker}, \citenamefont {Jorgensen},\ and\ \citenamefont
  {Olsen}}]{helgaker}%
  \BibitemOpen
  \bibfield  {author} {\bibinfo {author} {\bibfnamefont {T.}~\bibnamefont
  {Helgaker}}, \bibinfo {author} {\bibfnamefont {P.}~\bibnamefont {Jorgensen}},
  \ and\ \bibinfo {author} {\bibfnamefont {J.}~\bibnamefont {Olsen}},\
  }\href@noop {} {\emph {\bibinfo {title} {Molecular electronic-structure
  theory}}}\ (\bibinfo  {publisher} {John Wiley \& Sons},\ \bibinfo {year}
  {2014})\BibitemShut {NoStop}%
\bibitem [{\citenamefont {Ceperley}\ and\ \citenamefont
  {Alder}(1984)}]{releasenode}%
  \BibitemOpen
  \bibfield  {author} {\bibinfo {author} {\bibfnamefont {D.~M.}\ \bibnamefont
  {Ceperley}}\ and\ \bibinfo {author} {\bibfnamefont {B.~J.}\ \bibnamefont
  {Alder}},\ }\href@noop {} {\bibfield  {journal} {\bibinfo  {journal} {J.
  Chem. Phys.}\ }\textbf {\bibinfo {volume} {81}},\ \bibinfo {pages} {5833}
  (\bibinfo {year} {1984})}\BibitemShut {NoStop}%
\bibitem [{\citenamefont {Ceperley}(1995)}]{ceprev}%
  \BibitemOpen
  \bibfield  {author} {\bibinfo {author} {\bibfnamefont {D.~M.}\ \bibnamefont
  {Ceperley}},\ }\href@noop {} {\bibfield  {journal} {\bibinfo  {journal} {Rev.
  Mod. Phys.}\ }\textbf {\bibinfo {volume} {67}},\ \bibinfo {pages} {279}
  (\bibinfo {year} {1995})}\BibitemShut {NoStop}%
\bibitem [{\citenamefont {Booth}\ \emph {et~al.}(2009)\citenamefont {Booth},
  \citenamefont {Thom},\ and\ \citenamefont {Alavi}}]{alavi}%
  \BibitemOpen
  \bibfield  {author} {\bibinfo {author} {\bibfnamefont {G.~H.}\ \bibnamefont
  {Booth}}, \bibinfo {author} {\bibfnamefont {A.~J.~W.}\ \bibnamefont {Thom}},
  \ and\ \bibinfo {author} {\bibfnamefont {A.}~\bibnamefont {Alavi}},\
  }\href@noop {} {\bibfield  {journal} {\bibinfo  {journal} {J. Chem. Phys.}\
  }\textbf {\bibinfo {volume} {131}},\ \bibinfo {pages} {054106} (\bibinfo
  {year} {2009})}\BibitemShut {NoStop}%
\bibitem [{\citenamefont {Sorella}\ \emph {et~al.}(1989)\citenamefont
  {Sorella}, \citenamefont {Baroni}, \citenamefont {Car},\ and\ \citenamefont
  {Parrinello}}]{sorella}%
  \BibitemOpen
  \bibfield  {author} {\bibinfo {author} {\bibfnamefont {S.}~\bibnamefont
  {Sorella}}, \bibinfo {author} {\bibfnamefont {S.}~\bibnamefont {Baroni}},
  \bibinfo {author} {\bibfnamefont {R.}~\bibnamefont {Car}}, \ and\ \bibinfo
  {author} {\bibfnamefont {M.}~\bibnamefont {Parrinello}},\ }\href@noop {}
  {\bibfield  {journal} {\bibinfo  {journal} {EPL (Europhysics Letters)}\
  }\textbf {\bibinfo {volume} {8}},\ \bibinfo {pages} {663} (\bibinfo {year}
  {1989})}\BibitemShut {NoStop}%
\bibitem [{\citenamefont {Rossi}\ \emph {et~al.}(2009)\citenamefont {Rossi},
  \citenamefont {Nava}, \citenamefont {Reatto},\ and\ \citenamefont
  {Galli}}]{pigsnava}%
  \BibitemOpen
  \bibfield  {author} {\bibinfo {author} {\bibfnamefont {M.}~\bibnamefont
  {Rossi}}, \bibinfo {author} {\bibfnamefont {M.}~\bibnamefont {Nava}},
  \bibinfo {author} {\bibfnamefont {L.}~\bibnamefont {Reatto}}, \ and\ \bibinfo
  {author} {\bibfnamefont {D.~E.}\ \bibnamefont {Galli}},\ }\href@noop {}
  {\bibfield  {journal} {\bibinfo  {journal} {J. Chem. Phys.}\ }\textbf
  {\bibinfo {volume} {131}},\ \bibinfo {pages} {154108} (\bibinfo {year}
  {2009})}\BibitemShut {NoStop}%
\bibitem [{\citenamefont {Troyer}\ and\ \citenamefont
  {Wiese}(2005)}]{signproblem}%
  \BibitemOpen
  \bibfield  {author} {\bibinfo {author} {\bibfnamefont {M.}~\bibnamefont
  {Troyer}}\ and\ \bibinfo {author} {\bibfnamefont {U.}~\bibnamefont {Wiese}},\
  }\href@noop {} {\bibfield  {journal} {\bibinfo  {journal} {Phys. Rev. Lett.}\
  }\textbf {\bibinfo {volume} {94}},\ \bibinfo {pages} {170201} (\bibinfo
  {year} {2005})}\BibitemShut {NoStop}%
\bibitem [{\citenamefont {Ceperley}\ and\ \citenamefont
  {Mitas}(2007)}]{cep_book1}%
  \BibitemOpen
  \bibfield  {author} {\bibinfo {author} {\bibfnamefont {D.~M.}\ \bibnamefont
  {Ceperley}}\ and\ \bibinfo {author} {\bibfnamefont {L.}~\bibnamefont
  {Mitas}},\ }\enquote {\bibinfo {title} {Quantum {M}onte {C}arlo {M}ethods in
  {C}hemistry},}\ in\ \href {\doibase 10.1002/9780470141526.ch1} {\emph
  {\bibinfo {booktitle} {Advances in Chemical Physics}}}\ (\bibinfo
  {publisher} {John Wiley \& Sons, Inc.},\ \bibinfo {year} {2007})\ pp.\
  \bibinfo {pages} {1--38}\BibitemShut {NoStop}%
\bibitem [{\citenamefont {Feynman}\ and\ \citenamefont
  {Hibbs}(1965)}]{feynhibbs}%
  \BibitemOpen
  \bibfield  {author} {\bibinfo {author} {\bibfnamefont {R.~P.}\ \bibnamefont
  {Feynman}}\ and\ \bibinfo {author} {\bibfnamefont {A.~R.}\ \bibnamefont
  {Hibbs}},\ }\href@noop {} {\emph {\bibinfo {title} {Quantum Mechanics and
  Path Integrals}}}\ (\bibinfo  {publisher} {McGraw-Hill companies, New York},\
  \bibinfo {year} {1965})\BibitemShut {NoStop}%
\bibitem [{\citenamefont {Parrinello}\ and\ \citenamefont
  {Rahman}(1982)}]{kalos}%
  \BibitemOpen
  \bibfield  {author} {\bibinfo {author} {\bibfnamefont {M.}~\bibnamefont
  {Parrinello}}\ and\ \bibinfo {author} {\bibfnamefont {A.}~\bibnamefont
  {Rahman}},\ }in\ \href@noop {} {\emph {\bibinfo {booktitle} {Monte Carlo
  Methods in Quantum Problems}}},\ \bibinfo {editor} {edited by\ \bibinfo
  {editor} {\bibfnamefont {M.}~\bibnamefont {Kalos}}}\ (\bibinfo  {publisher}
  {D. Reidel Publishing Company},\ \bibinfo {year} {1982})\ Chap.\ \bibinfo
  {chapter} {Study of an F Center in Molten KCl}, pp.\ \bibinfo {pages}
  {105--116}\BibitemShut {NoStop}%
\bibitem [{\citenamefont {Parrinello}\ and\ \citenamefont
  {Rahman}(1984)}]{fcenter}%
  \BibitemOpen
  \bibfield  {author} {\bibinfo {author} {\bibfnamefont {M.}~\bibnamefont
  {Parrinello}}\ and\ \bibinfo {author} {\bibfnamefont {A.}~\bibnamefont
  {Rahman}},\ }\href@noop {} {\bibfield  {journal} {\bibinfo  {journal} {J.
  Chem. Phys.}\ }\textbf {\bibinfo {volume} {80}},\ \bibinfo {pages} {860}
  (\bibinfo {year} {1984})}\BibitemShut {NoStop}%
\bibitem [{\citenamefont {Marx}\ and\ \citenamefont
  {Parrinello}(1996)}]{pimd_par}%
  \BibitemOpen
  \bibfield  {author} {\bibinfo {author} {\bibfnamefont {D.}~\bibnamefont
  {Marx}}\ and\ \bibinfo {author} {\bibfnamefont {M.}~\bibnamefont
  {Parrinello}},\ }\href@noop {} {\bibfield  {journal} {\bibinfo  {journal}
  {The Journal of Chemical Physics}\ }\textbf {\bibinfo {volume} {104}},\
  \bibinfo {pages} {4077} (\bibinfo {year} {1996})}\BibitemShut {NoStop}%
\bibitem [{\citenamefont {Berne}\ and\ \citenamefont
  {Thirumalai}(1986)}]{berne}%
  \BibitemOpen
  \bibfield  {author} {\bibinfo {author} {\bibfnamefont {B.~J.}\ \bibnamefont
  {Berne}}\ and\ \bibinfo {author} {\bibfnamefont {D.}~\bibnamefont
  {Thirumalai}},\ }\href@noop {} {\bibfield  {journal} {\bibinfo  {journal}
  {Ann. Rev. Phys. Chem.}\ }\textbf {\bibinfo {volume} {37}},\ \bibinfo {pages}
  {401} (\bibinfo {year} {1986})}\BibitemShut {NoStop}%
\bibitem [{\citenamefont {Wolynes}\ and\ \citenamefont
  {Chandler}(1987)}]{wolchand}%
  \BibitemOpen
  \bibfield  {author} {\bibinfo {author} {\bibfnamefont {P.~G.}\ \bibnamefont
  {Wolynes}}\ and\ \bibinfo {author} {\bibfnamefont {D.}~\bibnamefont
  {Chandler}},\ }\href@noop {} {\bibfield  {journal} {\bibinfo  {journal} {J.
  Chem. Phys.}\ }\textbf {\bibinfo {volume} {74}},\ \bibinfo {pages} {4078}
  (\bibinfo {year} {1987})}\BibitemShut {NoStop}%
\bibitem [{\citenamefont {Laio}\ and\ \citenamefont
  {Parrinello}(2002)}]{meta_pnas}%
  \BibitemOpen
  \bibfield  {author} {\bibinfo {author} {\bibfnamefont {A.}~\bibnamefont
  {Laio}}\ and\ \bibinfo {author} {\bibfnamefont {M.}~\bibnamefont
  {Parrinello}},\ }\href@noop {} {\bibfield  {journal} {\bibinfo  {journal}
  {Proc. Natl. Acad Sci}\ }\textbf {\bibinfo {volume} {99}},\ \bibinfo {pages}
  {12562} (\bibinfo {year} {2002})}\BibitemShut {NoStop}%
\bibitem [{\citenamefont {Barducci}\ \emph {et~al.}(2008)\citenamefont
  {Barducci}, \citenamefont {Bussi},\ and\ \citenamefont {Parrinello}}]{wtm}%
  \BibitemOpen
  \bibfield  {author} {\bibinfo {author} {\bibfnamefont {A.}~\bibnamefont
  {Barducci}}, \bibinfo {author} {\bibfnamefont {G.}~\bibnamefont {Bussi}}, \
  and\ \bibinfo {author} {\bibfnamefont {M.}~\bibnamefont {Parrinello}},\
  }\href@noop {} {\bibfield  {journal} {\bibinfo  {journal} {Phys. Rev. Lett.}\
  }\textbf {\bibinfo {volume} {100}},\ \bibinfo {pages} {020603} (\bibinfo
  {year} {2008})}\BibitemShut {NoStop}%
\bibitem [{\citenamefont {Valsson}\ \emph {et~al.}(2016)\citenamefont
  {Valsson}, \citenamefont {Tiwary},\ and\ \citenamefont
  {Parrinello}}]{valsson}%
  \BibitemOpen
  \bibfield  {author} {\bibinfo {author} {\bibfnamefont {O.}~\bibnamefont
  {Valsson}}, \bibinfo {author} {\bibfnamefont {P.}~\bibnamefont {Tiwary}}, \
  and\ \bibinfo {author} {\bibfnamefont {M.}~\bibnamefont {Parrinello}},\
  }\href@noop {} {\bibfield  {journal} {\bibinfo  {journal} {Annual review of
  physical chemistry}\ }\textbf {\bibinfo {volume} {67}},\ \bibinfo {pages}
  {159} (\bibinfo {year} {2016})}\BibitemShut {NoStop}%
\bibitem [{\citenamefont {Ellenberger}\ \emph {et~al.}(2006)\citenamefont
  {Ellenberger}, \citenamefont {Ihn}, \citenamefont {Yannouleas}, \citenamefont
  {Landman}, \citenamefont {Ensslin}, \citenamefont {Driscoll},\ and\
  \citenamefont {Gossard}}]{qdot3}%
  \BibitemOpen
  \bibfield  {author} {\bibinfo {author} {\bibfnamefont {C.}~\bibnamefont
  {Ellenberger}}, \bibinfo {author} {\bibfnamefont {T.}~\bibnamefont {Ihn}},
  \bibinfo {author} {\bibfnamefont {C.}~\bibnamefont {Yannouleas}}, \bibinfo
  {author} {\bibfnamefont {U.}~\bibnamefont {Landman}}, \bibinfo {author}
  {\bibfnamefont {K.}~\bibnamefont {Ensslin}}, \bibinfo {author} {\bibfnamefont
  {D.}~\bibnamefont {Driscoll}}, \ and\ \bibinfo {author} {\bibfnamefont
  {A.}~\bibnamefont {Gossard}},\ }\href@noop {} {\bibfield  {journal} {\bibinfo
   {journal} {Phys. Rev. Lett.}\ }\textbf {\bibinfo {volume} {96}},\ \bibinfo
  {pages} {126806} (\bibinfo {year} {2006})}\BibitemShut {NoStop}%
\bibitem [{\citenamefont {Ceperley}(1996)}]{ceperley1995fermion}%
  \BibitemOpen
  \bibfield  {author} {\bibinfo {author} {\bibfnamefont {D.~M.}\ \bibnamefont
  {Ceperley}},\ }in\ \href@noop {} {\emph {\bibinfo {booktitle} {Monte Carlo
  and Molecular Dynamics of Condensed Matter Systems}}},\ Vol.~\bibinfo
  {volume} {49},\ \bibinfo {editor} {edited by\ \bibinfo {editor}
  {\bibfnamefont {K.}~\bibnamefont {Binder}}\ and\ \bibinfo {editor}
  {\bibfnamefont {G.}~\bibnamefont {Ciccotti}}}\ (\bibinfo  {publisher}
  {Compositori},\ \bibinfo {address} {Bologna, Italy},\ \bibinfo {year}
  {1996})\ Chap.\ \bibinfo {chapter} {Quantum {M}onte {C}arlo {M}ethods for
  {F}ermions}\BibitemShut {NoStop}%
\bibitem [{\citenamefont {Dama}\ \emph {et~al.}(2014)\citenamefont {Dama},
  \citenamefont {Parrinello},\ and\ \citenamefont {Voth}}]{WT_Dama}%
  \BibitemOpen
  \bibfield  {author} {\bibinfo {author} {\bibfnamefont {J.~F.}\ \bibnamefont
  {Dama}}, \bibinfo {author} {\bibfnamefont {M.}~\bibnamefont {Parrinello}}, \
  and\ \bibinfo {author} {\bibfnamefont {G.~A.}\ \bibnamefont {Voth}},\
  }\href@noop {} {\bibfield  {journal} {\bibinfo  {journal} {Phys. Rev. Lett.}\
  }\textbf {\bibinfo {volume} {112}},\ \bibinfo {pages} {240602} (\bibinfo
  {year} {2014})}\BibitemShut {NoStop}%
\bibitem [{\citenamefont {Tiwary}\ and\ \citenamefont
  {Parrinello}(2015)}]{reweighting}%
  \BibitemOpen
  \bibfield  {author} {\bibinfo {author} {\bibfnamefont {P.}~\bibnamefont
  {Tiwary}}\ and\ \bibinfo {author} {\bibfnamefont {M.}~\bibnamefont
  {Parrinello}},\ }\href@noop {} {\bibfield  {journal} {\bibinfo  {journal} {J.
  Phys. Chem. B}\ }\textbf {\bibinfo {volume} {119}},\ \bibinfo {pages} {736}
  (\bibinfo {year} {2015})}\BibitemShut {NoStop}%
\bibitem [{\citenamefont {Ceriotti}\ \emph {et~al.}(2009)\citenamefont
  {Ceriotti}, \citenamefont {Bussi},\ and\ \citenamefont {Parrinello}}]{GLE}%
  \BibitemOpen
  \bibfield  {author} {\bibinfo {author} {\bibfnamefont {M.}~\bibnamefont
  {Ceriotti}}, \bibinfo {author} {\bibfnamefont {G.}~\bibnamefont {Bussi}}, \
  and\ \bibinfo {author} {\bibfnamefont {M.}~\bibnamefont {Parrinello}},\
  }\href {\doibase 10.1103/PhysRevLett.102.020601} {\bibfield  {journal}
  {\bibinfo  {journal} {Phys. Rev. Lett.}\ }\textbf {\bibinfo {volume} {102}},\
  \bibinfo {pages} {020601} (\bibinfo {year} {2009})}\BibitemShut {NoStop}%
\bibitem [{\citenamefont {Herman}\ \emph {et~al.}(1982)\citenamefont {Herman},
  \citenamefont {Bruskin},\ and\ \citenamefont {Berne}}]{berne2}%
  \BibitemOpen
  \bibfield  {author} {\bibinfo {author} {\bibfnamefont {M.}~\bibnamefont
  {Herman}}, \bibinfo {author} {\bibfnamefont {E.}~\bibnamefont {Bruskin}}, \
  and\ \bibinfo {author} {\bibfnamefont {B.}~\bibnamefont {Berne}},\
  }\href@noop {} {\bibfield  {journal} {\bibinfo  {journal} {The Journal of
  Chemical Physics}\ }\textbf {\bibinfo {volume} {76}},\ \bibinfo {pages}
  {5150} (\bibinfo {year} {1982})}\BibitemShut {NoStop}%
\bibitem [{\citenamefont {Bennett}(1976)}]{bennett1}%
  \BibitemOpen
  \bibfield  {author} {\bibinfo {author} {\bibfnamefont {C.~H.}\ \bibnamefont
  {Bennett}},\ }\href@noop {} {\bibfield  {journal} {\bibinfo  {journal} {J.
  Comp. Phys.}\ }\textbf {\bibinfo {volume} {22}},\ \bibinfo {pages} {245}
  (\bibinfo {year} {1976})}\BibitemShut {NoStop}%
\bibitem [{\citenamefont {Ihn}(2015)}]{qdot1}%
  \BibitemOpen
  \bibfield  {author} {\bibinfo {author} {\bibfnamefont {T.}~\bibnamefont
  {Ihn}},\ }\href@noop {} {\emph {\bibinfo {title} {Semiconductor
  Nanostructures}}}\ (\bibinfo  {publisher} {Oxford University Press},\
  \bibinfo {year} {2015})\BibitemShut {NoStop}%
\bibitem [{\citenamefont {Reimann}\ and\ \citenamefont
  {Manninen}(2002)}]{qdot2}%
  \BibitemOpen
  \bibfield  {author} {\bibinfo {author} {\bibfnamefont {S.~M.}\ \bibnamefont
  {Reimann}}\ and\ \bibinfo {author} {\bibfnamefont {M.}~\bibnamefont
  {Manninen}},\ }\href@noop {} {\bibfield  {journal} {\bibinfo  {journal} {Rev.
  Mod. Phys.}\ }\textbf {\bibinfo {volume} {74}},\ \bibinfo {pages} {1283}
  (\bibinfo {year} {2002})}\BibitemShut {NoStop}%
\end{thebibliography}%

\end{document}